\documentstyle[12pt]{article}
\newcommand{\beq}{\begin{equation}}
\newcommand{\eeq}{\end{equation}}
\newcommand{\eq}[1]{(\ref{#1})}
\newcommand{\A}{{\bf a}}
\newcommand{\B}{{\bf b}}
\newcommand{\C}{{\bf c}}
\newcommand{\beqn}{\begin{eqnarray}}
\newcommand{\eeqn}{\end{eqnarray}}
\newcommand{\dst}{&\displaystyle}
\newcommand{\al}{\mbox{$Z\alpha$}}
\newcommand{\eps}{\mbox{$\varepsilon$}}
\newcommand{\Q}{\mbox{$\kappa$}}
\newcommand{\r}{\mbox{${\bf r}$}}
\newcommand{\k}{\mbox{${\bf k}$}}

\newcommand{\ro}{\mbox{\boldmath ${\rho}$\unboldmath}}

\newcommand{\e}{\mbox{${\bf e}$}}
\newcommand{\bi}[1]{\bibitem{#1}}
\newcommand{\fr}[2]{\frac{#1}{#2}}

\newcommand{\vu}{\mbox{${\bf u}$}}

\newcommand{\q}{\mbox{${\bf q}$}}

\newcommand{\vd}{\mbox{${\bf \Delta}$}}
\newcommand{\f}{\mbox{${\bf f}$}}
\begin{document}
\begin{titlepage}

\begin{center}
{\Large \bf Budker Institute of Nuclear Physics}
\end{center}

\vspace{1cm}

\begin{flushright}
{\bf Budker INP 97-85\\
October 22, 1997 }
\end{flushright}

\vspace{1.0cm}
\begin{center}{\Large \bf Large Coulomb corrections in
high-energy  photon splitting}\\ \vspace{1.0cm}

{\bf R.N. Lee, A.I. Milstein, V.M. Strakhovenko} \\
G.I. Budker Institute of Nuclear Physics,\\
630090 Novosibirsk, Russia\\

\vspace{4.0cm}
\end{center}

\begin{abstract}
The Coulomb corrections to the helicity amplitudes of high-energy
photon splitting are examined. The consideration is based on the
amplitudes obtained exactly in the parameter $\al$ within the
quasiclassical approach valid for small angles between all photon
momenta. We consider the case when the transverse momenta of both
final photons are  much larger than the electron mass.  It is shown
that at $\al\sim 1$ the Coulomb corrections essentially change the
result for the cross section as compared to the Born approximation.
The effect of screening is also taken into account.

\end{abstract}
\end{titlepage}

\section{\bf Introduction}

One of the most interesting nonlinear QED processes at high energy is
splitting of one photon into two in electric fields of atoms.
The total cross section of this process does not decrease with
increasing photon energy. First results on the observation of
high-energy photon splitting on atoms have been obtained recently in
the Budker Institute of Nuclear Physics \cite{Budker}.  Theoretically
this process has been investigated in \cite{Sh,CTP,B,JMO,S} only in
the lowest order in $\al$ (Born approximation), $Z|e|$ is the nucleus
charge, $\alpha =\, e^2/4\pi\, =1/137$ is the fine-structure
constant, $e$ is the electron charge.  Though the expressions
obtained in \cite{Sh,CTP} are rather cumbersome, some numerical
calculations based on the results of these papers have been carried
out in \cite{JMO,S} . Using the Weizs\"acker-Williams method
providing the logarithmic accuracy the cross section of the process
has been obtained in an essentially simpler form in  \cite{B}.  The
comparison of the exact cross section \cite{JMO} with the approximate
result \cite{B} has shown that the accuracy is better than $20\%$.
The Coulomb corrections to the cross section, which can essentially
modify the result as compared to that obtained in the Born
approximation, have been unknown up to now.

Recently we derived in \cite{LMS} the analytical expressions for
the high-energy photon-splitting amplitudes. The result was obtained
exactly in the parameter $\al$ at small angles $f_2$ and $f_3$
between the momenta $\k_2$ , $\k_3$ of the final photons and the
momentum $\k_1$ of the initial one. This region of angles gives
the main contribution to the total cross section of the process.
Small angles and high energies of photons allow one to use
the quasiclassical approach developed in \cite{MS,LM} at the
investigation of coherent photon scattering in a Coulomb
field (Delbr\"uck scattering). This approach gives the
transparent picture of the phenomenon and essentially simplifies the
calculation.

In the present paper we start from the analitycal results obtained
in \cite{LMS} and investigate the role of the Coulomb corrections
in the photon splitting process.  We restrict ourselves to the case
$|\k_{2\perp}|=\omega_2 f_2\gg m$, $|\k_{3\perp}|=\omega_3 f_3\gg
m$ ( $\omega_i=|\k_i|$, $m$ is the
electron mass) when the amplitudes can be essentially simplified.  A
pure Coulomb potential as well as the influence of screening is
considered.


\section{\bf Kinematics of the process}

Below we use the coordinate system with $z$-axis directed along
$\k_1$ so that $a_z=\A\k_1/\omega_1$ and
$\A_\perp=\A-a_z\k_1/\omega_1$ for an arbitrary vector $\A$.
 According to the
uncertainty relation the lifetime of the virtual electron-positron
pair is $\tau\sim |\r_2 -\r_1 | \sim
\omega_1/(m^2+\tilde{\Delta}^2)$, where
$\tilde{\Delta}=\mbox{max}(|\k_{2\perp}|,\,|\k_{3\perp}|)\ll\omega_1$.
The characteristic transverse distance
between the virtual particles can be estimated as
$(m^2+\tilde{\Delta}^2)^{-1/2}$, which is much smaller than the
length of the electron-positron loop. The characteristic impact
parameter is $\varrho\sim 1/\Delta$ , where ${\bf
\Delta}=\k_2+\k_3-\k_1$ is the momentum transfer. At small
$\k_{2\perp}$ and $\k_{3\perp}$ ($f_{2,3}\ll 1$) we have
\beq\label{Delta}
\vd^2=\, (\k_{2\perp}+\k_{3\perp}\,)^2+\frac{1}{4}
\left(\frac{\k_{2\perp}\,^2}{\omega_2}+
\frac{\k_{3\perp}\,^2}{\omega_3}\right)^2\, .
\eeq
The characteristic angular momentum is $l\sim \omega/\Delta \gg 1$ ,
and the quasiclassical approximation can be applied.

Let us discuss a screened Coulomb potential.
In the Thomas-Fermi model the screening radius is $r_{c}\sim
(m\alpha)^{-1}Z^{-1/3}$ .
If $R\ll 1/\Delta \ll r_{c}$ ($R$ is the nucleus radius), then the
screening is inessential and the amplitude coincides with that in the
pure Coulomb field. At $1/\Delta \sim r_{c}$ the screening
should be taken into account. Obviously, the impact parameters
$\varrho\gg r_{c}$ do not contribute to the total cross section. Due to
this fact we shall concentrate ourselves on the momentum transfer
region corresponding to the impact parameter $\varrho\leq r_{c}$.
If $\k_{2\perp}\,^2/\omega_2+\k_{3\perp}\,^2/\omega_3 \ll r_{c}^{-1}$
, then it follows from \eq{Delta} that the condition $\varrho\leq r_{c}$
holds only when $|{\bf \Delta}_{\perp}|=|\k_{2\perp}+\k_{3\perp}|\geq
r_{c}^{-1}$.  Thus, the main contribution to the amplitude
is given by the region of momentum transfer ${\bf \Delta}_{\perp}$,
restricted from below. In addition, at $\omega/(m^2+\tilde{\Delta}^2)
\gg r_{c} $ the angles between $\k_{1,2,3}$ and $\r_{1,2,3}$ are
either small or close to $\pi$ , and corresponding expansions are
used in our calculations.

According to the Furry theorem the photon-splitting amplitude is an
odd function with respect to the parameter $\al$. In the Born
approximation the amplitude is proportional to the Fourier transform
of the Coulomb potential ($\sim \al/\Delta^2$). Therefore, the
region of very small momentum transfers $\Delta \sim r_{c}^{-1}$ is
essential, and screening should be taken into account. In next
orders of perturbation theory with respect to the parameter $\al$
( Coulomb corrections ) the integral over all momenta corresponding
to the external field should be taken provided that their sum is
equal to ${\bf \Delta}$.  Therefore, even at $\Delta \sim r_{c}^{-1}$
each momentum is not small and the screening can be neglected. In the
Born approximation the screening can be taken into account by
multiplying the amplitude by the factor $[1-F(\Delta^2)]$, where
$F(\Delta^2)$ is the atomic electron form factor. Thus, to find the
photon-splitting amplitude in a screened Coulomb field it is
sufficient to solve the problem in a pure Coulomb field.


\section{Amplitudes}

It is convenient to perform
the calculations in terms of the helicity amplitudes \\
$M_{\lambda_1\lambda_2\lambda_3}(\k_1,\k_2,\k_3)$.
The longitudinal components of the polarization vectors $\e_i$ can be
eliminated owing to the relation $\e_i\k_i=0$ which leads to
$e_z=-\e_\perp\k_\perp/\omega$. After that within a small-angle
approximation one can neglect the difference between vectors
$(\e_{2,3})_\perp$ and the polarization vectors of photons
propagating along $\k_1$ and having the same helicities.
Therefore, the amplitudes $M_{\lambda_1\lambda_2\lambda_3}(\k_1,\k_2,\k_3)$
are expressed in terms of the polarization vectors $\e$ and $\e^*$
corresponding to positive and negative helicities, respectively.
It is sufficient to calculate three
amplitudes, for instance, $M_{+--}(\k_1,\k_2,\k_3)$ ,
$M_{+++}(\k_1,\k_2,\k_3)$ and $M_{++-}(\k_1,\k_2,\k_3)$ . The rest
amplitudes can be obtained by substitutions:

$$
M_{+-+}(\k_1,\k_2,\k_3)=M_{++-}(\k_1,\k_3,\k_2)\, , \quad
M_{-\lambda_2\lambda_3}(\k_1,\k_2,\k_3)=
M_{+\Lambda_2\Lambda_3}(\k_1,\k_2,\k_3)\,
(\e\leftrightarrow \e^*)\, ,
$$

where $\Lambda$ denotes the helicity opposite to $\lambda$
Let us introduce two-dimensional vectors
$\f_2=\k_{2\perp}/\omega_2$, $\f_3=\k_{3\perp}/\omega_3$
($|\f_{2,3}|\ll 1$) and $\f_{23}=\f_2-\f_3$.  We represent the
amplitudes obtained in \cite{LMS} for  $|\k_{2\perp}|\gg m$,
$|\k_{3\perp}|\gg m$ in the following form

\beqn\label{FINAL}
\dst
M=\fr{8{e^3}}{\pi^2\omega_1\omega_2\omega_3\vd^2}
\int{d\q}\,({\bf T}\mbox{\boldmath ${\nabla}$\unboldmath}_{\bf q})\,
{\mbox{Im} }\,\left(\fr{|\q+\vd|}{|\q-\vd|}\right)^{2iZ\alpha}
\quad ;\\
\dst
{}\nonumber\\
\dst
{\bf T}_{+--}=\e\int\limits_{0}^{\omega_2}\! d\eps\,
\fr{\Q_2\,(\e^*,\Q_2\f_2-\vd)}{4(\e^*\f_3)(\e^*\f_{23})(\e^*\A)}\,+\,
{{\omega_2\leftrightarrow\omega_3}\choose{\f_2\leftrightarrow\f_3}}\quad ;
\nonumber\\
\dst
{}\nonumber\\
\dst
{\bf T}_{+++}=\e^*\,\int\limits_{0}^{\omega_2}\! d\eps \,
\Biggl[\fr{\omega_3\eps\Q_2}{\omega_2{\cal D}_3}\left(
\fr{\omega_3}{2}+\fr{(\e^*\f_3)}{(\e^*\A)}(\Q_2^2+\Q_3^2)\right)+
\fr{(\Q_2^2+\Q_3^2)}{8(\e^*\A)(\e\f_{23})}\Biggr]\, +\,
{{\omega_2\leftrightarrow\omega_3}\choose{\f_2\leftrightarrow\f_3}}\, ;
\nonumber  \\
\dst
{}\nonumber\\
\dst
{\bf T}_{++-}=\e\,\int\limits_{0}^{\omega_2}\! d\eps \,
\Biggl[\fr{\omega_3\Q_2\Q_3}{\omega_1{\cal D}_1}
\left(
\fr{\Q_2-\eps}{2}-\fr{(\e\f_{23})}{(\e\A)}(\Q_2^2+\eps^2)
\right)+\fr{(\Q_2^2+\eps^2)}{8(\e^*\f_3)(\e\A)}\Biggr]+\nonumber  \\
\dst
\e\,\omega_2\,\int\limits_{-\omega_3}^{0}\! d\eps \,
\Q_3\Biggl[
\fr{\Q_3(\Q_2^2+\eps^2)}{(\e^*\B)}\left(
\fr{(\e^*\f_{23})}{\omega_1{\cal D}_1}+
\fr{(\e^*\f_{2})}{\omega_3{\cal D}_2}
\right)
-\fr{\eps\Q_3+\Q_2^2}{2\omega_1{\cal D}_1} +
\fr{\eps^2-\Q_2\Q_3}{2\omega_3{\cal D}_2}\Biggr]\, .
\nonumber
\eeqn
where $\q$ is a two-dimensional vector lying in the plane
perpendicular to $\k_1$, $\vd$ denotes $\vd_\perp$.

In eq. \eq{FINAL} the following notation is used:
\beqn\label{abc} \dst
\Q_2=\omega_2-\eps\, , \quad \Q_3=\omega_3+\eps\,,\nonumber\\
\dst
{\cal D}_1=\fr{\omega_2\Q_3\A^2-\omega_3\Q_2\B^2}{\omega_1\eps}-i0\,\,
,\,\,
{\cal D}_2=\fr{\omega_2\Q_3\tilde{\C}^2-\omega_1\eps\B^2}{\omega_3\Q_2}
\,\, ,\,\,
{\cal D}_3=\fr{\omega_1\eps\A^2+\omega_3\Q_2\C^2}{\omega_2\Q_3}
\,\, ,
 \\
\dst
\A=\q-\vd+2\Q_2\f_2\quad , \quad \B=\q+\vd-2\Q_3\f_3\quad ,
\nonumber
\\ \dst
\C=\q+\vd-2\eps\f_{23}\quad ,\quad
\tilde{\C}=\q-\vd+2\eps\f_{23}\quad .\nonumber
\eeqn

Note that vectors $\e$ and
$\e^*$ appeared in denominators in \eq{FINAL} owing to the
use of the relation $2(\e\A)(\e^*\A)=\A^2$.

As it was mentioned above, the photon-splitting amplitude obtained
in the Born approximation \cite{Sh,CTP} for arbitrary energies and
momentum transfers is rather cumbersome. It is interesting
to obtain the Born amplitude from \eq{FINAL}. In this approximation
\beqn\dst (\e\mbox{\boldmath
${\nabla}$\unboldmath}_{\bf q})\, {\mbox{Im}
}\,\left(\fr{|\q+\vd|}{|\q-\vd|}\right)^{2iZ\alpha} \longrightarrow
\al\,\left[\fr{1}{(\e^*,\q+\vd)}-\fr{1}{(\e^*,\q-\vd)}\right]\nonumber\, .
\eeqn
It is convenient to rewrite the quantities ${\cal D}_{1-3}$ in
\eq{abc} as

\begin{eqnarray}\label{newD}
\dst
{\cal D}_1=\left(\q+\fr{\Q_2-\Q_3}{\omega_1}\,\vd
\right)^2-\fr{4\omega_2\omega_3\Q_2\Q_3}{\omega_1^2}\f_{23}^2-i0\quad
, \\
\dst
{\cal D}_2=\left(\q-\fr{\Q_3+\epsilon}{\omega_3}\,\vd
\right)^2-\fr{4\omega_1\omega_2\Q_3\epsilon}{\omega_3^2}\f_{2}^2\quad
,\nonumber\\
\dst
{\cal D}_3=\left(\q+\fr{\Q_2-\epsilon}{\omega_2}\,\vd
\right)^2+\fr{4\omega_1\omega_3\Q_2\epsilon}{\omega_2^2}\f_{3}^2\quad
.\nonumber
\end{eqnarray}
After that we shift the variable of integration $\q$ in each term so
that the quantities ${\cal D}_{1-3}$ become independent of the angle
$\phi$  of the vector $\q$. For instance, in the terms containing
${\cal D}_1$ we make the substitution $\q\rightarrow
\q-\fr{\Q_2-\Q_3}{\omega_1}\,\vd$. After passing to the
variable $z=\exp(i\phi)$ we can easily take the corresponding contour
integral. Taking also
the integrals with respect to $|\q|$ and $\epsilon$, we get
for the Born amplitudes
\beqn\label{Born}
\dst
M_{+--}=\fr{2i\al e^3
(\f_2\times\f_3)_z}{\pi\vd^2(\e^*\f_2) (\e^*\f_3)
(\e^*\f_{23})}\quad,\\ \dst \nonumber\\ \dst \nonumber\\ \dst
M_{+++}=\fr{2(\al) e^3\omega_1}{\pi\vd^2
(\e\f_{23})^2\omega_2\omega_3}\Biggl\{(\e\vd)\Biggl[1+\fr{(\e\f_2)+(\e\f_3)}
{(\e\f_{23})}\ln({a_2\over
a_3})+
\nonumber\\
\dst
\nonumber\\
\dst
\fr{(\e\f_2)^2+(\e\f_3)^2}{(\e\f_{23})^2}\left({\pi^2\over 6}
+{1\over 2} \ln^2({a_2\over
a_3})+\mbox{Li}_2(1-a_2)+\mbox{Li}_2(1-a_3)\right)\Biggr]+\nonumber\\
\dst
\nonumber\\
\dst
{1\over (\e\vd)}\left[\omega_3^2(\e\f_3)^2\fr{a_2}{1-a_2}\left(
1+{a_2\,\ln(a_2)\over
1-a_2}\right)+\omega_2^2(\e\f_2)^2\fr{a_3}{1-a_3}
\left( 1+{a_3\,\ln(a_3)\over 1-a_3}\right)
\right]+\nonumber\\
\dst
\nonumber\\
\dst
\fr{2(\e\f_2)(\e\f_3)}{(\e\f_{23})}\left[\omega_3{a_2\,
\ln(a_2)\over 1-a_2}-
\omega_2{a_3\,\ln(a_3)\over 1-a_3}\right]\Biggr\}\quad ,\nonumber \\
\nonumber\\
\dst
\nonumber\\
\dst
M_{++-}=\fr{2(\al) e^3\omega_2}{\pi\vd^2
(\e^*\f_{3})^2\omega_1\omega_3}\Biggl\{(\e^*\vd)\Biggl[1-
\fr{(\e^*\f_2)+(\e^*\f_{23})}{(\e^*\f_3)}\ln({-a_1\over a_2})
+
\nonumber\\
\dst
\nonumber\\
\dst
\fr{(\e^*\f_2)^2+(\e^*\f_{23})^2}{(\e^*\f_3)^2}\left({\pi^2\over 6}
+{1\over 2} \ln^2({-a_1\over
a_2})+\mbox{Li}_2(1-a_2)+\mbox{Li}_2(1+a_1)\right)\Biggr]+\nonumber\\
\dst
\nonumber\\
\dst
{1\over (\e^*\vd)}\left[\omega_3^2(\e^*\f_{23})^2\fr{a_2}{1-a_2}\left(
1+{a_2\,\ln(a_2)\over
1-a_2}\right)-\omega_1^2(\e^*\f_2)^2\fr{a_1}{1+a_1}
\left( 1-{a_1\,\ln(-a_1)\over 1+a_1}\right)
\right]+\nonumber\\
\dst
\nonumber\\
\dst
\fr{2(\e^*\f_2)(\e^*\f_{23})}{(\e^*\f_3)}\left[
\omega_1{a_1\,\ln(-a_1)\over 1+a_1}
-\omega_3{a_2\,\ln(a_2)\over 1-a_2}\right]\Biggr\}\quad ,\nonumber \\
\nonumber
\eeqn
where
$$
a_1=\fr{\vd^2}{\omega_2\omega_3\f_{23}^2}\quad,\quad
a_2=\fr{\vd^2}{\omega_1\omega_2\f_2^2}\quad,\quad
a_3=\fr{\vd^2}{\omega_1\omega_3\f_3^2}\quad,\quad
\mbox{Li}_2(x)=-\int\limits_0^x\fr{dt}{t}\ln(1-t)\quad.
$$

It follows from \eq{abc}  that $\ln(-a_1)$ should be interpreted
as $\ln(-a_1+i0)=\ln(a_1)+i\pi$. Besides,
$$
\mbox{Li}_2(1+a_1)=\mbox{Li}_2(1+a_1-i0)=
\fr{\pi^2}{6}-\ln(1+a_1)
[\ln(a_1)+i\pi]-\mbox{Li}_2(-a_1)
$$
The result \eq{Born} is obtained for $|\vd_\perp|\gg
|\vd_z|$.  One can show that it remains valid in the case
$|\vd_\perp|\sim |\vd_z|$ if the expression \eq{Delta} for
$\vd^2$ is used in \eq{Born}. Actually, in eq. \eq{Born} the
difference between $\vd^2$ and $\vd_\perp^2$ is essential only in the
overall factor $1/\vd^2$.  For a screened Coulomb potential the
amplitudes \eq{Born} should be multiplied by the atomic form factor
$(1-F(\Delta^2))$.  For the case of Moli\`ere potential \cite{M} it
reads
\beq\label{FF}
1-F(\Delta^2)=\Delta^2\sum_{i=1}^{3}\,\fr{\alpha_{i}}{\Delta^2+\beta_{i}^2}
\, ,
\eeq
where
\beqn
\label{coef}
\dst
\alpha_{1}=0.1\quad , \quad \alpha_{2}=0.55 \quad,\quad
\alpha_{3}=0.35 \quad, \quad \beta_{i}=\beta_0 b_i \quad,
\\
\dst \nonumber
\\
\dst \nonumber
b_{1}=6 \quad, \quad b_{2}=1.2 \quad , \quad b_{3}=0.3 \quad ,
\quad \beta_0=\, mZ^{1/3}/121\quad .
\eeqn
Remind that the representation
\eq{Born} is valid when $|\k_{2\perp}|\, , \,|\k_{3\perp}|\gg m$.

Let us consider the asymptotics of the amplitudes \eq{FINAL} at
$|\vd_\perp|\ll |\ro|$, where $\vd_\perp=\omega_2\f_2+\omega_3\f_3$
and $\ro=(\omega_2\f_2-\omega_3\f_3)/2$. It is this region of
variables which gives the main contribution to the cross section in
the Weizs\"acker-Williams approximation.
To get this asymptotics we multiply ${\bf T}$ in \eq{FINAL} by
$$
1=\vartheta(q_0^2-\q^2)+\vartheta(\q^2-q_0^2) \quad,
$$
where $|\vd|\ll q_0\ll |\ro|$. Then, for the term in \eq{FINAL}
proportional to $\vartheta(q_0^2-\q^2)$ one can put $\q=0$ and
$\vd=0$ in ${\bf T}$ and integrate by parts over $\q$. After that,
using the relation $\mbox{\boldmath${\nabla}$\unboldmath}_{\bf q}
\vartheta(q_0^2-\q^2)= -2\q\, \delta(q_0^2-\q^2)$ one can easily
take the integral over $\q$ since at $|\q|=q_0\gg |\vd|$ one has
$$
{\mbox{Im} }\,\left(\fr{|\q+\vd|}{|\q-\vd|}\right)^{2iZ\alpha}\approx
4\al\,\fr{\q\vd}{\q^2} \quad .
$$
As a result, in the region $|\q|<q_0$ the term proportional
to $\al$ is independent of $q_0$ and the terms of next orders in
$\al$ are small in the parameter $|\vd|/q_0$.

For the term proportional to $\vartheta(\q^2-q_0^2)$ we get
$$
\mbox{\boldmath ${\nabla}$\unboldmath}_{\bf q}
{\mbox{Im} }\,\left(\fr{|\q+\vd|}{|\q-\vd|}\right)^{2iZ\alpha}\approx
4\al\,\fr{\q^2\vd-2\q(\q\vd)}{|\q|^4}\quad .
$$
We put $\vd=0$ in $\bf T$ and perform  the integration first over
the angles of $\q$ and then over $|\q|$.
As a result, the main in $q_0/|\ro|$ contribution is independent of
$q_0$ and proportional to \al. Taking  the sum of the contributions
from these two regions and performing the integration over the energy
$\eps$, we get
\beqn\label{zero} \dst
M_{+--}=\fr{4iN(\e\ro)^3}{\ro^4} (\vd\times\ro)_z\quad,\quad
N=\fr{4\al{e^3}\omega_2\omega_3}{\pi\omega_1\vd^2\ro^2}\,;
\\
\dst
{}\nonumber\\
\dst
M_{+++}=N
\Biggl[\e^*\vd+
2(\e\vd)\fr{(\e^*\ro)^2}{\ro^2}\left(1+
\fr{\omega_2-\omega_3}{\omega_1}\ln\fr{\omega_3}{\omega_2}+
\fr{\omega_2^2+\omega_3^2}{2\omega_1^2}(\ln^2\fr{\omega_3}{\omega_2}+\pi^2)\right)
\Biggr]\nonumber\\
\dst
{}\nonumber\\
\dst
M_{++-}=N
\Biggl[\e\vd+2(\e^*\vd)\fr{(\e\ro)^2}{\ro^2}\left(1+
\fr{\omega_1+\omega_3}{\omega_2}(\ln\fr{\omega_3}{\omega_1}+i\pi)+
\fr{\omega_1^2+\omega_3^2}{2\omega_2^2}(\ln^2\fr{\omega_3}{\omega_1}+
2i\pi\ln\fr{\omega_3}{\omega_1})\right)
\Biggr]\, .\nonumber
\eeqn
In the small-angle approximation ($|\f_2|,|\f_3|\ll 1$) the cross
section of the process reads:

\beq \label{cross1}
d\sigma=\fr{\omega_1^2}{2^8 \pi^5}|M|^2\, x(1-x)dx \, d\f_2\, d\f_3\,
\quad,
\eeq
where $x=\omega_2/\omega_1$, so that $\omega_3=\omega_1(1-x)$.
 In terms of the variables $\ro$ and
$\vd$ the cross section has the form
\beq  \label{cross2}
d\sigma=|M|^2\,\fr{d\vd\, d\ro\, dx}{2^8 \pi^5 \omega_1^2
\,x(1-x)}\quad,
\eeq
Substituting \eq{zero} into
\eq{cross2} and performing the elementary integration over the
angles of vectors $\vd$ and $\ro$, we come to the expression
\beq\label{sa} d\sigma=\fr{4Z^2
\alpha^5}{\pi^2}\fr{d\rho^2\,d\Delta^2\,dx}{\rho^4\Delta^2}\,g(x)\quad ,
\eeq
where the function $g(x)$ for different polarizations has the form
\beqn\label{g}
\dst
g_{+--}(x)=x(1-x)\quad ,\\
\dst
g_{+++}(x)=\fr{1}{2}x(1-x)\Biggl[1+
\Biggl(1+(2x-1)\ln\left(\fr{1-x}{x}\right)+\nonumber\\
\dst
\fr{x^2+(1-x)^2}{2}\left(\ln^2\left(\fr{1-x}{x}\right)+\pi^2\right)
\Biggr)^2\,\Biggr]\quad ,
\nonumber\\
\dst
g_{++-}(x)=\fr{1}{2}x(1-x)\Biggl[1+\Biggl|1+
(2/x-1)(\ln(1-x)+i\pi)+\nonumber\\
\dst
\fr{1+(1-x)^2}{2x^2}(\ln^2(1-x)+
2i\pi\ln(1-x))\Biggr|^2\,\Biggr]\quad
,\nonumber\\
\dst
g_{+-+}(x)=g_{++-}(1-x)\quad .\nonumber
\eeqn
Formulae \eq{sa} and \eq{g} are in agreement with the corresponding
results of \cite{B}, obtained in the Weizs\"acker-Williams
approximation.  However, this approach does not allow to obtain the
amplitudes \eq{zero} themselves.  The large logarithm
appears after the integration of \eq{sa} over $\Delta^2$ from
$\Delta_{min}^2$ up to $\rho^2$ where $\Delta_{min}\sim r_c^{-1}$ for
the screened Coulomb potential and $\Delta_{min}\sim
\rho^{2}/\omega_1$ for the pure Coulomb case. It is interesting to
compare the contributions of different helicity amplitudes to the
cross section at $\Delta\rightarrow 0$. In Fig. 1 the function $g(x)$
is shown for different helicities as well as the quantity
\beq\label{gmean}
\bar{g}(x)=g_{+--}(x)+g_{+++}(x)+g_{++-}(x)+g_{+-+}(x)\, ,
\eeq
which corresponds to the summation over the final photon
polarizations. It is seen that $\bar{g}(x)$ has a wide plateau.

The Coulomb corrections to the photon-splitting amplitude at
$\Delta\rightarrow 0$ are small compared to the Born term \eq{zero}.
We consider the asymptotics of the Coulomb corrections at
$\Delta\rightarrow 0$ in the next Section.

\section{Coulomb corrections}

To analyze the Coulomb corrections we make the further
transformation of the expression \eq{FINAL}. Let us multiply right
side of \eq{FINAL} by
$$
1\equiv\int\limits_{-1}^{1}dy\,\delta \left(y-\fr{2\q\vd}{\q^2+\vd^2}
\right)=
(\q^2+\vd^2)\int\limits_{-1}^{1}\fr{dy}{|y|}\,
\delta((\q-\vd /y)^2 -\vd^2(1/y^2-1))\quad ,
$$
and make the substitution $\q\rightarrow \q+\vd/y$.
After that the integral over $|\q|$ becomes trivial, and the integral
over the angle of $\q$ can be easily taken by means of the residue
technique. Finally, we get

\beqn\label{FINAL1} \dst
M=\fr{4e^3\al}{\pi\omega_1\omega_2\omega_3\vd^2}
\int\limits_{-1}^{1}\fr{dy\,\,\mbox{sign}\,y}{1-y^2}\,\left[\mbox{Re}
\left(\fr{1+y}{1-y}\right)^{iZ\alpha}\right]\, R\, ; \\
\dst
{}\nonumber\\
\dst
 R_{+--}=\fr{q^2}{(\e^*\f_3)(\e^*\f_{23})}
\int\limits_{0}^{\omega_2}\! d\eps\,\Q_2\,(\e^*,\Q_2\f_2-\vd)
\,\fr{\vartheta(\r^2-q^2)}{(\e^*\r)^2}\,+\,
{{\omega_2\leftrightarrow\omega_3}\choose{\f_2\leftrightarrow\f_3}}\quad ;
\nonumber\\
\dst
{}\nonumber\\
\dst
R_{+++}=\int\limits_{0}^{\omega_2}\! d\eps \,
\Biggl[\fr{\omega_3\eps\Q_2}{\omega_2}\left(
\fr{s}{\sqrt{s^2-4q^2\vu^2}}-1\right)\left(\fr{\omega_3}{(\e\vu)}-
\fr{8(\e^*\f_3)(\Q_2^2+\Q_3^2)}{s-\sqrt{s^2-4q^2\vu^2}-
4(\e^*\r)(\e\vu)}\right)+
\nonumber  \\
\dst
{}\nonumber\\
\dst
\vartheta(q^2-\r^2)\fr{(\Q_2^2+\Q_3^2)\Q_2(1+1/y)(\e^*\f_{2})(\e\vd)}
{(\e\f_{23})[\eps(\e^*\r)(\e\f_{23})-\Q_2(1+1/y)(\e\vd)(\e^*\f_2)]}
\Biggr]\,+
\,{{\omega_2\leftrightarrow\omega_3}
\choose{\f_2\leftrightarrow\f_3}}\quad;
\nonumber  \\
\dst
{}\nonumber\\
\dst
R_{++-}=\int\limits_{0}^{\omega_2}\! d\eps \,
\Biggl[\fr{\omega_3\Q_2\Q_3}{\omega_1}
\left(\fr{is_1}{\sqrt{4q^2\vu^2_1-s^2_1}}-1\right)\left(
\fr{\Q_2-\eps}{(\e^*\vu_1)}+\fr{8(\e\f_{23})(\Q_2^2+\eps^2)}
{s_1+i\sqrt{4q^2\vu^2_1-s^2_1}-4(\e\r)(\e^*\vu_1)}\right)+
\nonumber  \\
\dst
{}\nonumber\\
\dst
\vartheta(q^2-\r^2)\fr{(\Q_2^2+\eps^2)\Q_2(1+1/y)(\e\f_{2})(\e^*\vd)}
{(\e^*\f_{3})[\Q_3(\e\r)(\e^*\f_{3})-\Q_2(1+1/y)(\e^*\vd)(\e\f_2)]}
\Biggr]-\nonumber  \\
\dst
{}\nonumber\\
\dst
\omega_2\int\limits_{-\omega_3}^{0}\! d\eps \,
\Q_3\Biggl[\fr{(\e\vd)}{\omega_1(\e^*\vd)}
\left(\fr{is_1}{\sqrt{4q^2\vu^2_1-s^2_1}}-1\right)
\left(\fr{\eps\Q_3+\Q_2^2}{(\e\vu_1)}+\fr{8(\e^*\f_{23})\Q_3(\Q_2^2+\eps^2)}
{s_1-i\sqrt{4q^2\vu^2_1-s^2_1}-4(\e^*\r_1)(\e\vu_1)}\right)+
\nonumber  \\
\dst
{}\nonumber  \\
\dst
\fr{(\e\vd)}{\omega_3(\e^*\vd)}
\left(\fr{s_2}{\sqrt{s^2_2-4q^2\vu^2_2}}-1\right)
\left(\fr{\Q_2\Q_3-\eps^2}{(\e\vu_2)}+\fr{8(\e^*\f_{2})\Q_3(\Q_2^2+\eps^2)}
{s_2+\sqrt{s^2_2-4q^2\vu^2_2}-4(\e^*\r_1)(\e\vu_2)}\right)+
\nonumber  \\
\dst
{}\nonumber  \\
\dst
\vartheta(\r^2_1-q^2)\fr{q^2(\Q_2^2+\eps^2)}{2\omega_2(\e^*\r_1)}
\Biggl(\fr{1}{\Q_2(\e^*\r_1)(\e\f_{2})-\Q_3(1/y-1)(\e\vd)(\e^*\f_3)}+
\nonumber  \\
\dst
{}\nonumber  \\
\dst
\fr{1}{\eps(\e^*\r_1)(\e\f_{23})-\Q_3(1/y-1)(\e\vd)(\e^*\f_3)}
\Biggr)\Biggr]\, .
\nonumber
\eeqn
Here we use the following notation:
\beqn\label{FINALadd}
\dst
\vu=\vd\left(\fr{1}{y}-1+\fr{2\Q_2}{\omega_2}\right)\, ,\quad
\vu_1=\vd\left(\fr{1}{y}-1+\fr{2\Q_2}{\omega_1}\right)\, ,\quad
\vu_2=\vd\left(\fr{1}{y}-1-\fr{2\eps}{\omega_3}\right)\, ,\,\nonumber \\
\dst
{}\nonumber  \\
\dst
q^2=\vd^2(1/y^2-1)\, ,\quad\r=\vd(1/y-1)+2\Q_2\f_2\, ,\quad
\r_1=\vd(1/y+1)-2\Q_3\f_3\, ,\\
\dst
{}\nonumber  \\
\dst
s=\vu^2+q^2+\fr{4\omega_1\omega_3\Q_2\eps}{\omega_2^2}\f_3^2\, ,\,
s_1=\vu_1^2+q^2-\fr{4\omega_2\omega_3\Q_2\Q_3}{\omega_1^2}\f_{23}^2\,-i0\,
,\,
s_2=\vu_2^2+q^2-\fr{4\omega_1\omega_2\Q_3\eps}{\omega_3^2}\f_2^2\, .
\nonumber
\eeqn
Since the function $R$ in \eq{FINAL1} is independent of the parameter
$\al$ the Coulomb corrections $M^{(c)}$ can be obtained
from \eq{FINAL1} by the substitution
$$
\mbox{Re}\left(\fr{1+y}{1-y}\right)^{iZ\alpha}\rightarrow
\mbox{Re}\left(\fr{1+y}{1-y}\right)^{iZ\alpha}-1
$$
The asymptotics of $M^{(c)}$ at $\Delta\rightarrow 0$ depends on the
photon helicities. The most simple way to get this asymptotics is
to start directly from \eq{FINAL}.
For $M_{+--}^{(c)}$ the main contribution is determined by the
region $q\sim\rho\gg\Delta$ (remind that
$\ro=(\omega_2\f_2-\omega_3\f_3)/2$), while for $M_{+++}^{(c)}$ and
$M_{++-}^{(c)}$ it comes from the region $q\sim\Delta\ll\rho$. After
the corresponding expansion and integration over the energy we get at
$\Delta\rightarrow 0$

\beqn\label{FINAS} \dst
M_{+--}^{(c)}=-\fr{2e^3(\al)^3\omega_2\omega_3\,(\e^*\vd)}
{\pi\omega_1(\e^*\ro)^4}\ln^2\frac{\rho}{\Delta}\quad ;\\
{}\dst
\nonumber\\
\dst
M_{+++}^{(c)}=-\fr{ie^3\al\omega_2\omega_3}{2\pi\omega_1\rho^2(\e\ro)
(\e^*\vd)}
\int\frac{d\q}{(\e,\q-\vd)}\left[
\mbox{Re}\,\left(\fr{|\q+\vd|}{|\q-\vd|}\right)^{2iZ\alpha}-1\right]\,
\mbox{sign}[(\q-\vd)\times\ro]_z\quad ;\nonumber\\
\dst
{}\nonumber\\
\dst
M_{++-}^{(c)}=-\fr{e^3\al\omega_2^2\omega_3}{2\pi^2\omega_1^2
\rho^2(\e^*\ro)(\e\vd)}\int\frac{d\q}{(\e^*,\q-\vd)}\left[
\mbox{Re}\,\left(\fr{|\q+\vd|}{|\q-\vd|}\right)^{2iZ\alpha}-1\right]\,\times
\nonumber\\
\dst
\left[\ln\fr{|\q-\vd|}{\Delta}+
i\mbox{arg}\fr{\e(\q-\vd)}{\e\ro}\right]\, .\nonumber
\eeqn
It follows from \eq{FINAS} that in this limiting case the Coulomb
correction $M_{+--}^{(c)}$ is small, while $M_{+++}^{(c)}$ and
$M_{++-}^{(c)}$ depend only on the direction of vector $\vd$, but not
on its module (it becomes obvious after the substitution
$\q\rightarrow \q \,\Delta$). We discuss the role of Coulomb
corrections in the next Section.

\section{Cross section}

As it was suggested in \cite{MW}, to overcome the problems of
background in the measurement of photon splitting one has to register
the events with $|\f_{2,3}|\geq f_0$ where $f_0 \ll 1$ is determined
by the experimental conditions. Let us consider the cross section
integrated over $\f_3$ for $|\f_3|>f_0$. It is interesting to compare
the exact ( in $\al$ ) cross section $d\sigma/dx\,d\f_2$ with that
obtained in the Born approximation ($d\sigma_{B}/dx\,d\f_2$) and also
with the cross section in the Weizs\"acker-Williams approximation
($d\sigma_{W}/dx\,d\f_2$). Remind that $d\sigma_{W}/dx\,d\f_2$ is in
fact the Born cross section calculated within logarithmic accuracy.
Large logarithm
corresponds to the contribution of the region $\Delta\ll
\rho=|\omega_2\f_2-\omega_3\f_3)/2|$, where $f_3\approx xf_2/(1-x)$.
Taking the integral over $\Delta^2$ in eq.  \eq{sa} from
$\Delta^2_{min}$ up to $\Delta^2_{eff}$, where (see \cite{B})
$$
\Delta^2_{min}=\Delta^2_{z}=(\omega_1 f_2^2x/2(1-x))^2\, ,
\, \Delta^2_{eff}=\rho^2=(\omega_1 x f_2)^2\, ,
$$
and summing over the final photon polarizations
we get for a pure Coulomb potential
\beqn\label{wwc}\dst
\fr{d\sigma_{W}}{dx\,d\f_2}=\fr{8Z^2\alpha^5 }
{\pi^3\omega_1^2}\,\fr{\bar{g}(x)}{x^2f_2^4}\,
\ln\left(\fr{2 (1-x)}{f_2}\right)\,\vartheta(\fr{x}{1-x}f_2-f_0) \, .
\eeqn
For the case of a screened Coulomb potential the approximate cross
section is
\beqn\label{wwcs}\dst
\fr{d\sigma_{W}}{dx\,d\f_2}=\fr{4Z^2\alpha^5 }
{\pi^3\omega_1^2}\,\fr{\bar{g}(x)}{x^2f_2^4}\,
\left[2\ln
\left(\fr{\omega_1xf_2}{\beta_0}\right)+\gamma
\right]\,\vartheta(\fr{x}{1-x}f_2-f_0)\, .
\eeqn
The function $\gamma$ in eq. \eq{wwcs} is
\beqn\label{gamma}\dst
\gamma=1-\sum\limits_{i=1}^3\,\alpha_i^2 (\ln a_i +1)
-2\sum\limits_{i>j}\,\alpha_i\alpha_j
\fr{a_i\ln a_i-a_j\ln a_j}{a_i-a_j}\, ,\,
a_i=b_i^2+\Delta^2_{min}/\beta_0^2
\eeqn
and the coefficients $\alpha_i$, $b_i$ and $\beta_0$ are defined in
eq. \eq{coef}. If $\Delta^2_{min}/\beta_0^2\gg 1$ then
$\gamma=-\ln(\Delta^2_{min}/\beta_0^2)$ and eq. \eq{wwcs} turns to
eq. \eq{wwc}.  If $\Delta^2_{min}/\beta_0^2\ll 1$ then
$\gamma=-0.158$.

For the case of a pure Coulomb potential the
dependence of $\sigma_0^{-1} d\sigma/dxd\f_2$ on $f_2/f_0$ is shown
in Figs. 2-4 for $f_0=10^{-3}$, $Z=92$ and different values of
$x$,
$$
\sigma_0=\fr{4Z^2\alpha^5}{\pi^3\omega_1^2 f_0^4}\quad .
$$

In these figures the solid curves represent the exact cross sections,
the dashed ones are Born results ($d\sigma_{B}/dx\,d\f_2$)
and the dash-dotted are obtained in the Weizs\"acker-Williams
approximation. At $x=0.7$ (Fig. 2) the difference between
$d\sigma_{B}/dx\,d\f_2$ and $d\sigma_{W}/dx\,d\f_2$ is small. At
$x=0.3$ (Fig. 3) $d\sigma_{B}/dx\,d\f_2$ differs noticeably from
$d\sigma_{W}/dx\,d\f_2$. Note that within a good accuracy the cross
section $d\sigma_{B}/dx$ at $x=0.3$ agrees with that obtained from
eq.  \eq{wwc} after the integration over $\f_2$.  It should be so,
since $d\sigma / dx$ is invariant with respect to the substitution
$x\rightarrow 1-x$ and at $x=0.7$, as we pointed out above, the
approximate result \eq{wwc} is in accordance with the exact one.  At
$x=0.5$ a big difference between $d\sigma_{B}/dx\,d\f_2$ and
$d\sigma_{W}/dx\,d\f_2$  (see Fig. 4) in the region  $f_2\sim f_0$
can be explained as follows. The large logarithm appears as a
result of integration with respect to $\f_3$ over the range
$|(1-x)\f_3+x\f_2|\ll x f_2$. After the integration over the azimuth
angle $\varphi$ between vectors $\f_2$ and $-\f_3$ we should
integrate over $f_3$ from $f_0$ up to $x f_2/(1-x)$ and from $x
f_2/(1-x)$ to infinity. If $x f_2/(1-x) \approx f_0$ then the
contribution of the first region vanishes and the cross section
becomes approximately two times smaller (in accordance with Fig. 4).
In all cases the exact cross section at $\al\sim 1$ is
noticeably smaller than the Born one. The magnitude of this effect
depends on kinematics. For instance, at $x=0.3$ (Fig. 3) and $f_2 <
f_0(1-x)/x$ when $\Delta\sim \rho$ the exact cross section is several
times smaller than the Born one, while for $f_2 > f_0(1-x)/x$ the
difference of these cross sections is about $15\%$. For $x=0.5$ and
$0.7$ (Figs. 3,4) this difference is about $20\%$ and almost
independent of $f_2/f_0$. All indicated relations between
cross sections take place also when screening is taken into account.
Emphasize that at $\Delta\sim\rho$ and $\al\sim 1$ the exact cross
section differential over all variables ($d\sigma/dx\,d\f_2d\f_3$) is
much smaller than the Born one. In Fig. 5 this
differential cross section is shown for the case of a screened
Coulomb potential. The peak for azimuth angle $\phi=\pi$
corresponds to small $\vd$. There is a narrow notch at
$f_3=x f_2/(1-x)$ which corresponds to the condition $\vd_\perp=0$.
The width $\delta f_3$ of the notch is about
$\mbox{max}(\Delta_z/\omega_3,\beta_0/\omega_3)$.  For the
parameters used in Fig. 5, $\delta f_3$ is about $10^{-4}$.

Let us discuss now the cross section $d\sigma/dx$. In the
Weizs\"acker-Williams approximation for a pure Coulomb
potential the cross section $d\sigma_{W}/dx$ obtained from \eq{wwc}
is
\beqn\label{wwtot} \dst
\fr{d\sigma_{W}}{dx} = \pi
f_0^2\,\sigma_0\bar{g}(x)\,
\Biggl[\fr{\vartheta(x-1/2)}{x^2}\,
\left(2\ln \fr{2
(1-x)}{f_0}-1 \right)+(x \leftrightarrow 1-x)\Biggr]\, .
\eeqn

If $\omega_1f_0^2/(1-x)\ll\beta_0$, then the corresponding
cross section in a screened potential has the form
\beqn \dst
\fr{d\sigma_{W}}{dx} =\pi
f_0^2\,\sigma_0\bar{g}(x)\,\Biggl[\fr{\vartheta(x-1/2)}{x^2}\,
\left(2\ln
\fr{\omega_1xf_0}{\beta_0}+0.842\right)+(x \leftrightarrow
1-x)\Biggr]\, .
\eeqn

The inequality $\Delta\ll \rho$ which provides the applicability
of the Weizs\"acker-Williams approximation corresponds to a small
angle $\varphi$ between the vectors $\f_2$ and $-\f_3$ ( when the
vectors $\f_2$ and  $\f_3$ have almost opposite directions).  So, it
is interesting to consider the quantity
$d\sigma_B(\varphi_{max})/dx$ which is the Born cross section
integrated over the angle $\varphi$ from $-\varphi_{max}$ to
$\varphi_{max}$. In the case of a pure Coulomb potential the
dependence of $(\pi f_0^2 \sigma_0)^{-1} d\sigma_B(\varphi_{max})/dx$
on $\varphi_{max}$ is shown in Fig. 6 for different $x$ and
$f_0=10^{-3}$.  One can see that the cross section is saturated at
relatively large $\varphi_{max}$. The same conclusion is valid for
the case of a screened Coulomb potential.

Let us represent the exact cross section as a sum
$d\sigma/dx=d\sigma_{B}/dx+d\sigma_{C}/dx$.
As it was mentioned above, the region of small momentum transfers
$\Delta$ is not important for the Coulomb corrections $d\sigma_C/dx$.
Therefore, one can neglect the effects of screening in the
calculation of $d\sigma_{C}/dx$. The quantity $F=(\pi
f_0^2\sigma_0)^{-1}\,d\sigma_C/dx$ is independent of $f_0$ and
$\omega_1$ and is the function of $\al$ and $x$. The dependence of
$F$ on $x$ for different $Z$ is shown in Fig. 7. On can see that the
Coulomb corrections diminish the cross section of the process. To
realize the magnitude of this effect we plot in Fig. 8 the cross
sections $d\sigma/dx$, $d\sigma_B/dx$ and $d\sigma_W/dx$ for $Z=83$
and $f_0=10^{-3}$. It is seen that the Coulomb corrections are
important. This is the consequence of the fact that at $\al\sim 1$
the constant added to the logarithm, i.e.  the quantity
$x^2\,F/\bar{g}(x)$, is large enough in a wide range of $x$.
Although this quantity is independent of $f_0$ and $\omega_1$, the
relative contribution of the Coulomb corrections to the exact cross
section depends on these parameters since the large logarithm
contains them.

Due to the gauge invariance the cross section $d\sigma/dx$ should be
equal to zero at $x=1$ if the electron mass is not neglected. This is
not the case for massless particles as it has been noted in
\cite{KU}. That is why  the cross section
$d\sigma/dx$ calculated in zero-mass limit does not vanish at
$x\rightarrow 1$ (see Fig. 8) . When $x\rightarrow 1$
the main contribution to the cross section comes from the range of
angles $f_3\sim f_2/\sqrt{1-x}$ and $f_2\sim f_0$ , $\Delta\approx
\omega_2f_2$, where the Weizs\"acker-Williams approximation is not
applicable.  Since $f_3\ll 1$, the relation $1-x\gg f_0^2$
should be fulfilled. Besides, the condition $k_{3,\,\perp}\gg m$
means that $(1-x)\omega_1\,f_3\sim\sqrt{1-x}\omega_1\,f_0\gg m $ or
$1-x\gg (m/\omega_1f_0)^2$.

If we represent the amplitude $M$ of the
process as a sum of Born amplitude $M_B$ and Coulomb corrections
$M_C$, then $|M|^2=|M_B|^2+2\,\mbox{Re}(M_B^*M_C)+
|M_C|^2$. Taking into account in $M_C$ only the lowest in $\al$
term (proportional to $ (\al)^3$), we get $d\sigma^{(1)}_C/(\al)^4 =
c_1+c_2(\al)^2$ , where $c_1$ and $c_2$ are independent of $\al$ and
$c_2\, >\, 0$ . The dependence of $(\al)^{-2}(\pi
f_0^2\sigma_0)^{-1}\,d\sigma_C/dx$ on $\al$ is shown in Fig. 9
for different $x$. One can see that starting from $Z\sim 15$
the contribution of the next order Coulomb corrections essentially
modifies the behavior of $d\sigma_C$.

Thus, the Coulomb corrections are very important for the adequate
description of high-energy photon splitting and must be taken into
account at the comparison of theory and experiment.

We are grateful to V.S. Fadin and E.A. Kuraev for useful discussions.

\newpage \begin{center} Figure captions \end{center}



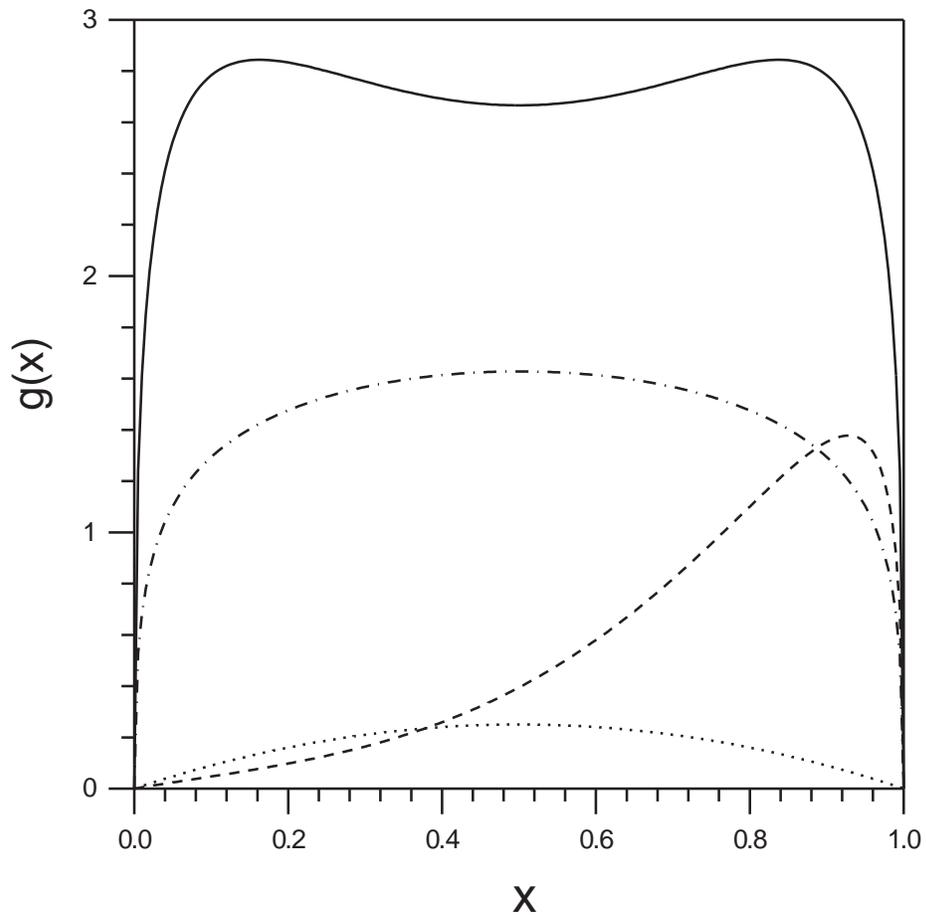
Fig. 1. Function $g(x)$ from eq. \eq{g} for different polarizations:
$g_{+--}(x)$ (dotted curve),$g_{++-}(x)$ (dashed curve),$g_{+++}(x)$
(dash-dotted curve), and $\bar{g}(x)$ (solid curve), eq. \eq{gmean}.

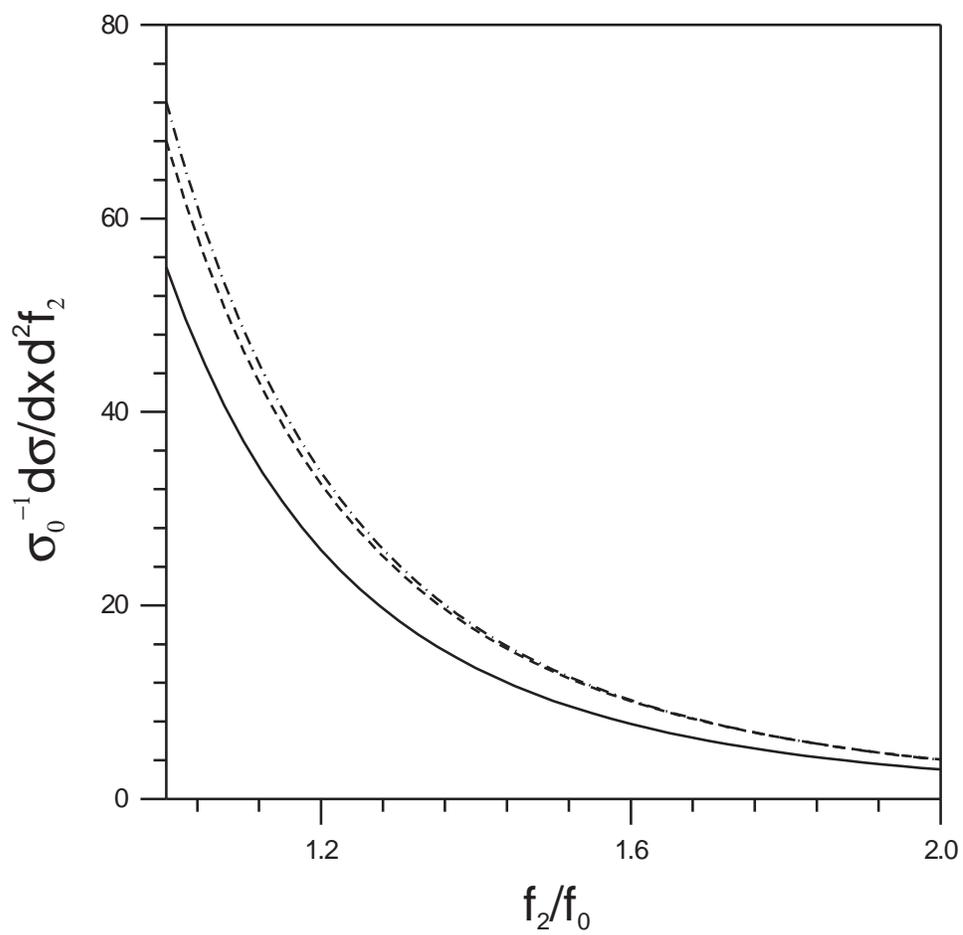
Fig. 2.  $\sigma_0^{-1} d\sigma/dxd\f_2$ versus $f_2/f_0$ for
a pure Coulomb potential, $f_0=10^{-3}$, $x=0.7$,
$Z=92$, $\sigma_0$ is given in the text. The dash-dotted curve
corresponds to the Weizs\"acker-Williams approximation, the dashed
curve gives the Born approximation, the solid curve is the exact
result.

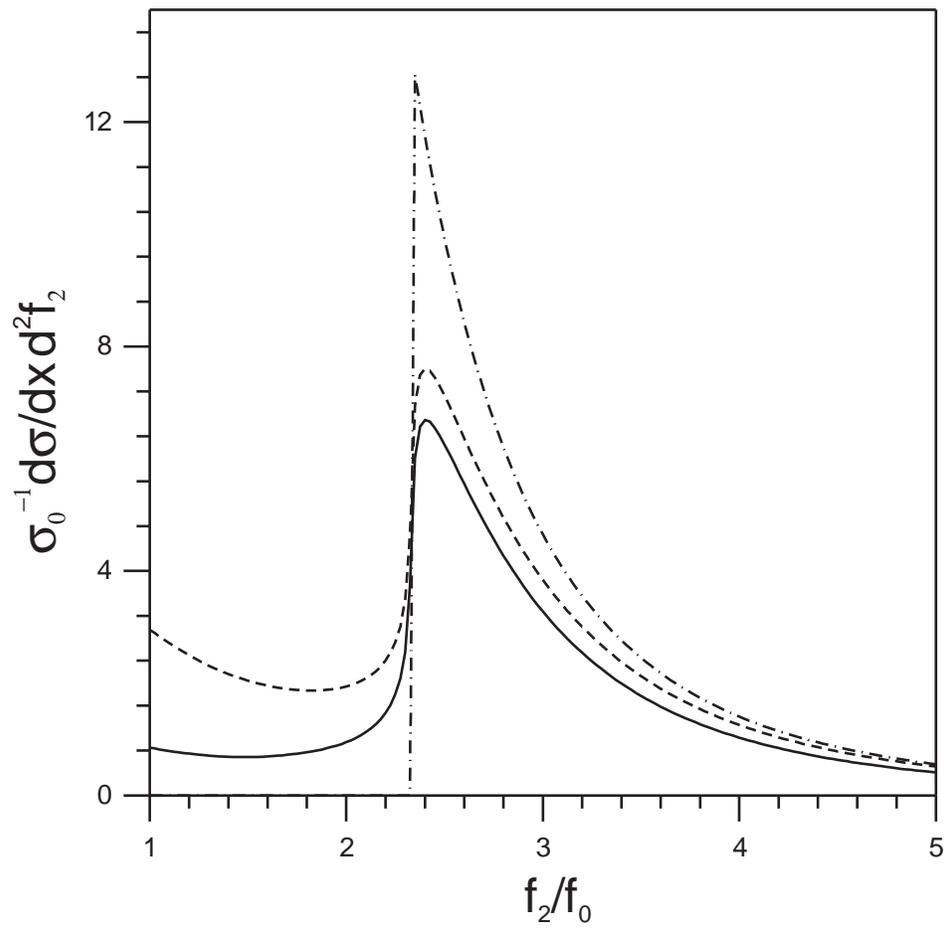
Fig. 3. Same as Fig. 4 but for $x=0.3$ .

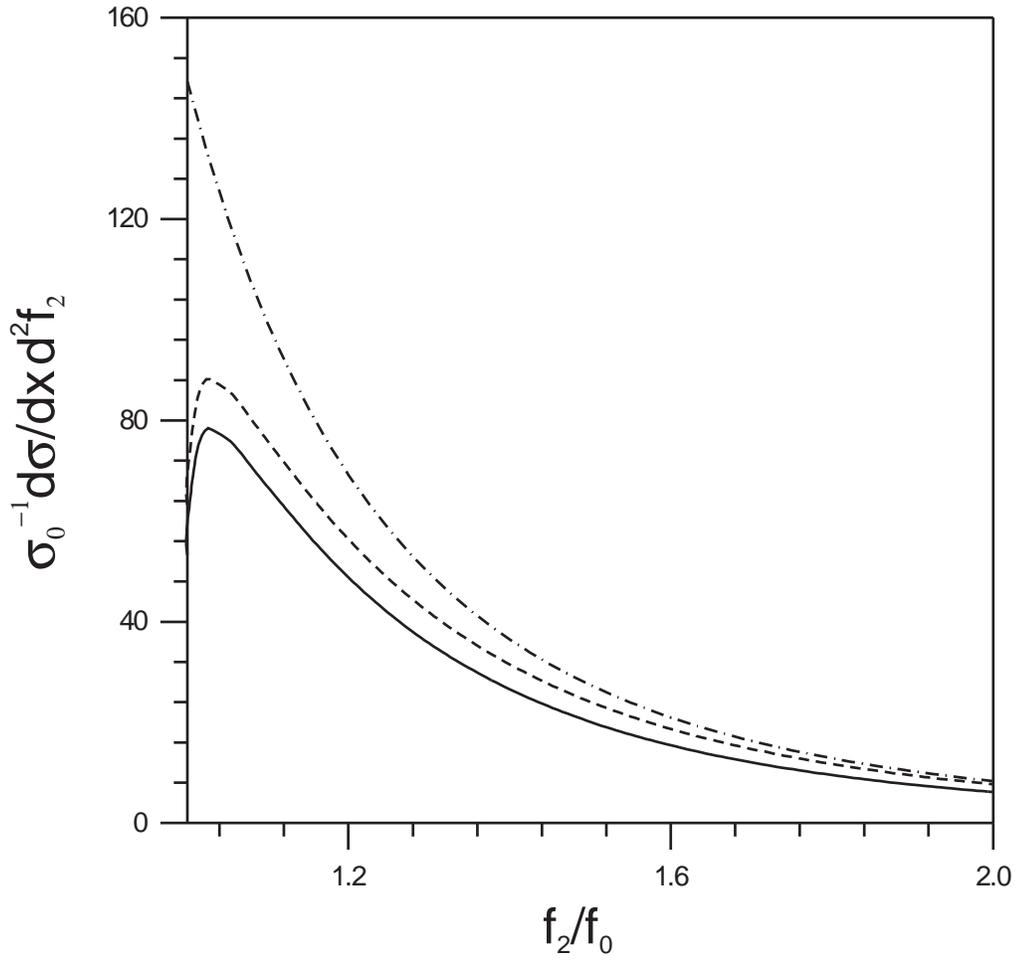
Fig. 4. Same as Fig. 4 but for $x=0.5$ .

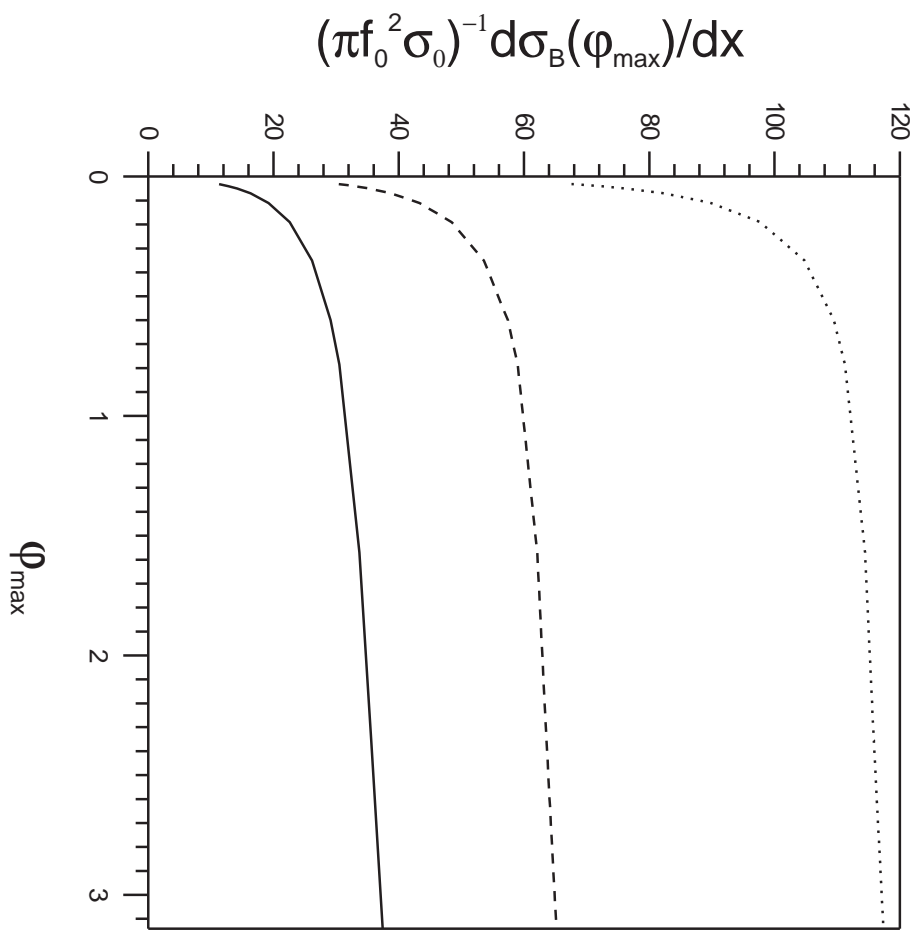
Fig. 5. Differential cross section $d\sigma/dxd\f_2d\f_3$ versus
$f_3$ in a screened Coulomb potential for different
azimuth angle $\phi$ between vectors $\f_2$ and $\f_3$; $Z=83$,
$x=0.7$, $\omega_1=1$GeV, $f_2=5$mrad.  The dashed curve (Born
approximation) and the solid curve (exact cross section) correspond to
$\phi=\pi$. The dash-dotted curve (Born approximation) and
the dotted curve (exact cross section) correspond to $\phi=\pi/2$.

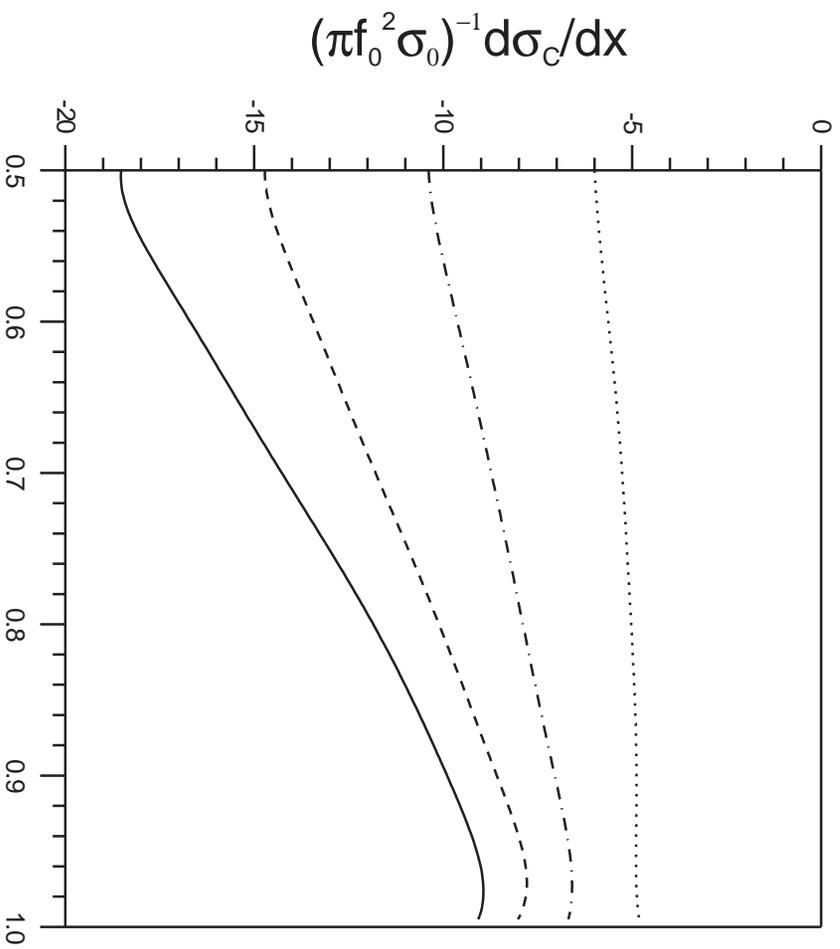
Fig. 6. The dependence of $(\pi f_0^2\sigma_0)^{-1}
d\sigma(\varphi_{max})/dx$ on $\varphi_{max}$ for a pure Coulomb
potential at different~$x$: $x=0.5$ (dotted curve), $x=0.7$ (dashed
curve), and $x=0.9$ (solid curve); $f_0=10^{-3}$.

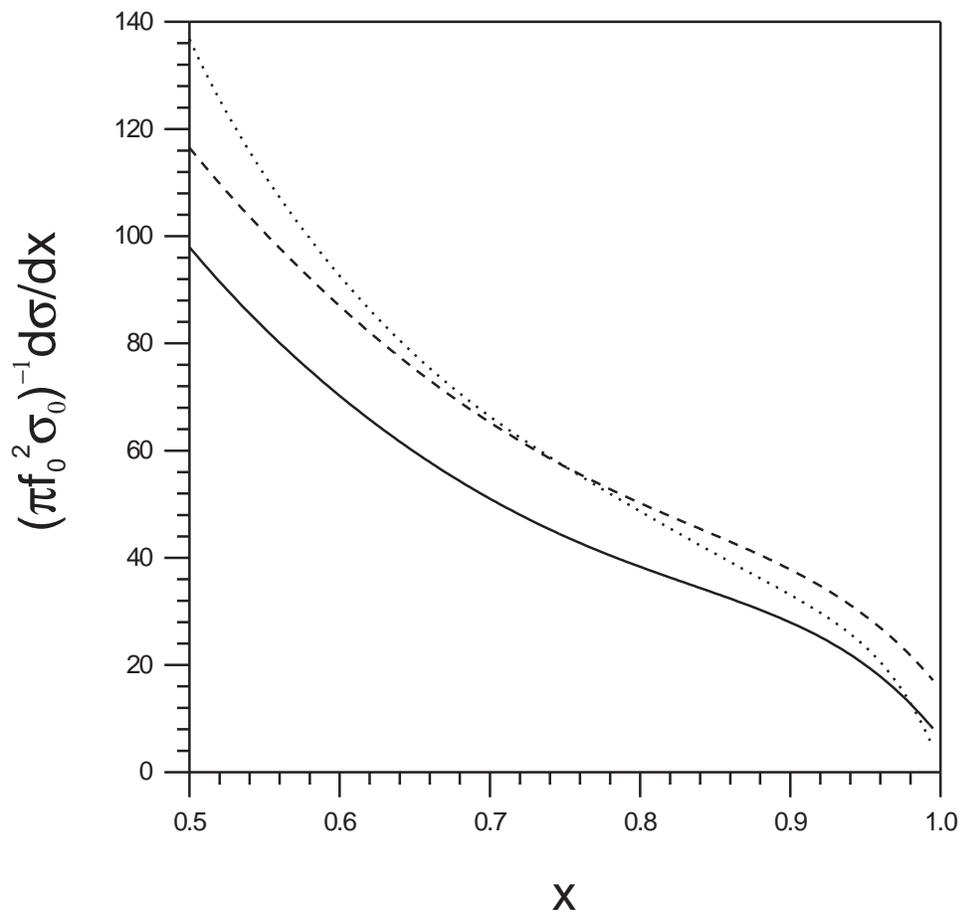
Fig. 7. The dependence of $(\pi f_0^2\sigma_0)^{-1}
d\sigma_C/dx$ on $x$ for Z=32 (dotted curve),
47 (dash-dotted curve), 64 (dashed curve) and 83 (solid curve).

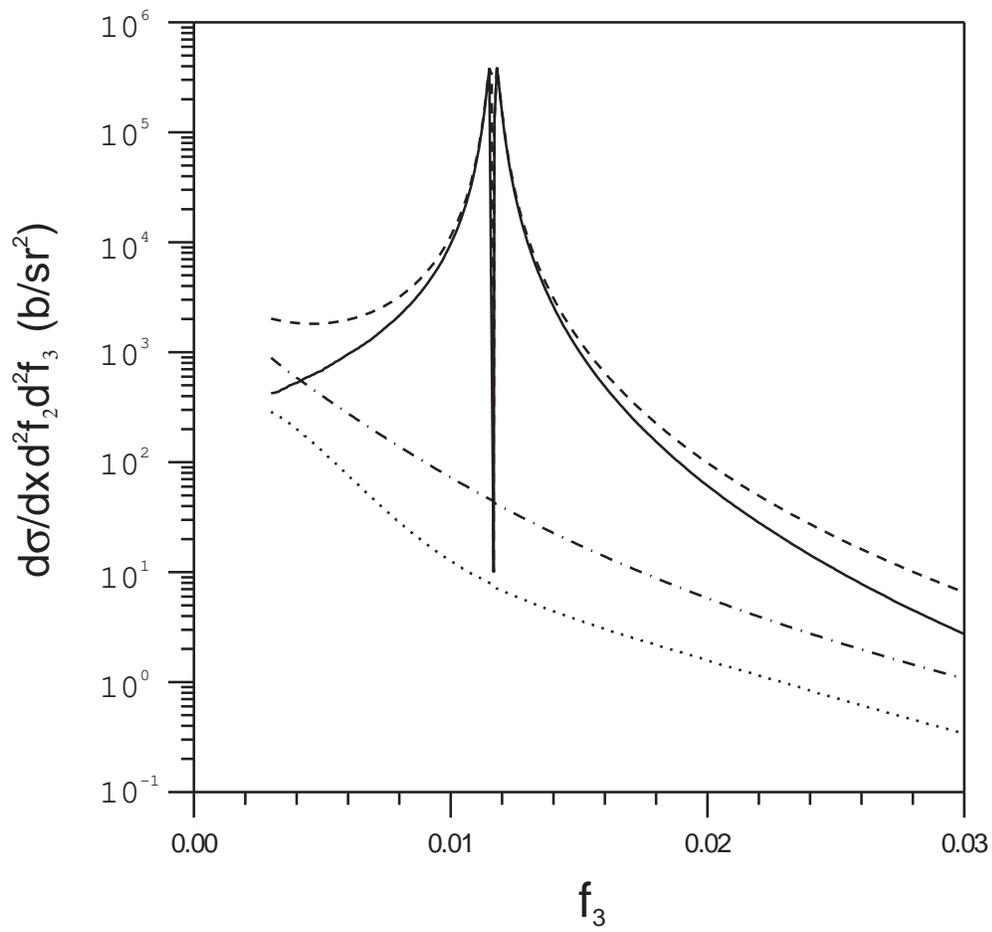
Fig. 8. The dependence of $(\pi f_0^2\sigma_0)^{-1}
d\sigma/dx$ on $x$ for Z=83. The dotted curve corresponds to the
Weizs\"acker-Williams approximation, the dashed curve gives the
Born approximation and the solid one shows the result exact in $\al$.

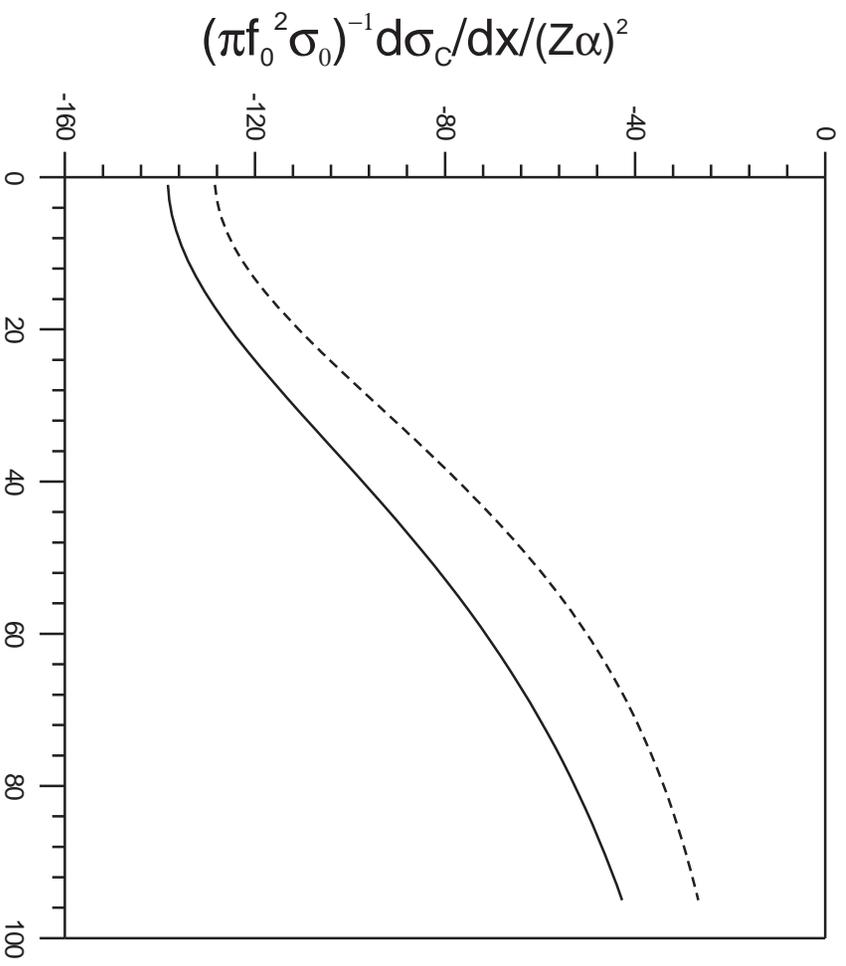
Fig. 9. The dependence of $(\pi f_0^2\sigma_0)^{-1}
d\sigma/dx/\al^2$ on $Z$ for x=0.5 (solid curve) and
x=0.8 (dashed curve).

\end{document}